
\input amstex
\documentstyle{amsppt}
\NoRunningHeads
\TagsOnRight
\Monograph
\magnification=1200
\document
$$\split\qquad\qquad\qquad\qquad\qquad\qquad\qquad\qquad\qquad\qquad\qquad
\qquad\qquad\qquad &\text{UWThPh-94-56}\\ &\ \text{November 1994}
\endsplit$$
\bigskip\bigskip
\bigskip
$$\gather\text{{\bf YANG-MILLS THEORY WITH THE PONTRYAGIN TERM}}\\
\text{{\bf ON MANIFOLDS WITH A BOUNDARY}}\endgather$$
$$\gather\text{Gerald Kelnhofer}\\\text{Institut f\"ur Theoretische Physik,
Universit\"at Wien}\\ \text{Boltzmanngasse 5, 1090 Vienna}\\ \text{Austria}
\endgather$$
\bigskip
\bigskip
\bigskip
\bigskip
\bigskip
\bigskip
\bigskip
\bigskip
\bigskip
\head Abstract\endhead
The $3+1$ dimensional Yang-Mills theory with
the Pontryagin term included is studied on manifolds with a boundary. Based
on the geometry of the universal bundle for Yang-Mills theory, the
symplectic structure of this model is exhibited. The topological type of
the quantization line bundles is shown to be determined by the torsion
elements in the cohomology of the gauge orbit space.
\par\newpage
{\bf 1. Introduction}
\bigskip
Gauge theories whose actions combine an ordinary Yang-Mills kinetic term and
a topological term possess an interesting mathematical structure.
Prominent examples are the Yang-Mills theory in odd dimensions including the
Chern-Simons term [1,2] -afterwards called CSYM theory- and
secondly the
Yang-Mills theory in even dimensions with the Pontryagin term added to
which we refer as PYM theory [3].
These topological terms introduce functional
abelian background fields in the corresponding configuration space,
whose geometrical structure can be traced back to
cohomological properties of the Yang-Mills orbit space
$\Cal M$ [4-6]: Depending on the space dimension and the gauge group, a
consistent
quantization of the CSYM theory imposes a quantization condition on the
coupling parameter which has its origin in the second integer
cohomology group of $\Cal M$ [5,6]. On manifolds with a
nontrivial boundary the generators of gauge transformations satisfy an
anomalous commutator algebra which is cohomologically
equivalent to the Faddeev-Mickelsson current algebra [2,7].
Recently, the
discussion of pure Chern-Simons theory on manifolds with boundaries has lead
to a detailled investigation on so-called edge states [8], which carry
representations of the Kac-Moody algebra.\par
On the other hand, it is well known [3] that a Pontryagin term
($\theta$-term) can be added in nonabelian gauge
theories due to instanton effects. This gives rise to a kind
of
Aharonov Bohm effect where the $\theta$ parameter is identified with the
magnetic flux associated with a vortex structure in the gauge orbit
space [4].
Topologically, this effect is related to the first integer cohomology group
of $\Cal M$ [6]. \par
In this paper we consider the PYM-theory on manifolds with a nontrivial
boundary. We restrict ourselves to the case of 4 dimensional manifolds
but the generalization to higher dimensions is straightforward. Since we aim
at a Hamiltonian description, we assume that the 4-manifold $N$ is of the
form $N=\Bbb R\times M$ and that $M$ has a nontrivial boundary
$\partial M$. The purpose of this paper is to clarify the symplectic
structure of the classical phase space and to study the corresponding
quantum theory
of this model in terms of the geometry of the underlying gauge orbit space.
The starting point is the classical action
$$S=\int _{\Bbb R\times M}tr(F_{\bar A}\wedge\bar\star F_{
\bar A})+\frac{\theta}{8\pi ^2}
\int _{\Bbb R\times M}tr(F_{\bar A}\wedge F_{\bar A}),\tag1$$
where $F_{\bar A}=d\bar A+\frac{1}{2}[\bar A,\bar A ]$ is the Yang-Mills
field strength, $\bar A$ is regarded as connection
on a principal bundle $\bar P$ over $N$
with a compact, connected, simple Lie group $G$ as structure group,
and $\bar\star$ is the Hodge operator induced by a fixed Riemannian
structure on $N$.
\par
It will be shown that the
symplectic data of the action (1) are related to certain secondary
characteristic classes of
the universal bundle for Yang-Mills theory which originally has been
introduced by
Atiyah and Singer [9] in order to study nonabelian anomalies.
{}From the mathematical viewpoint, the contribution of the boundary can be
expressed in terms of a new set of descent equations in the universal bundle.
These equations have previously been used in the analysis of covariant
anomalies in Yang-Mills theory [10]. \par
The classical reduced phase space is the cotangent bundle $T^{\ast}\Cal M$
of the gauge orbit space equipped with a symplectic structure which differs
from the canonical one by the pullback of a closed two form defined on
the gauge
orbit space associated with the Yang-Mills fields on the boundary manifold.
\par
Using the
framework of geometric quantization [11], it turns out
that quantization of the PYM model is not unique, if the corresponding
gauge group is disconnected. Topologically, the quantum line bundles are
related to the torsion elements in the integer
cohomology group of $\Cal M$. As a consequence, no quantization condition
must be imposed on the coupling parameter $\theta$. The physical states are
invariant under infinitesimal gauge transformations but in general they
acquire a phase under gauge transformations not belonging to the identity
component of the gauge group.
\bigskip\bigskip
{\bf 2. The geometrical setup}
\bigskip
In this section we shall prepare the necessary mathematical constructions
in order to study the classical and quantum theory of the model defined by
(1). It will be shown in the next section that a symplectic
formulation of this model is intimately related with the geometrical
structure of the universal bundle for Yang-Mills theory. \par
Let $M$ be a compact, connected 3-manifold with
boundary $\Sigma :=\partial M$ and let $P(M,G)$ be the principal fiber bundle
with structure group $G$ which is given by restricting $\bar P$ to
$M$. Associated to $P$ with respect of the adjoint action of $G$ on its
Lie algebra $\frak g$ is the adjoint bundle $adP:=P\times _G\frak g$. The
gauge group $\Cal G$ is defined to be the group of those vertical bundle
automorphisms of $P$ which act freely on the space $\Cal A$
of all connections on $P$. For $A\in\Cal A$ and $u\in\Cal G$ this action is
given by the pull-back $A\mapsto u^{\ast}A$. Let
$\Cal A(\Cal M,\pi _{\Cal A},\Cal G)$ be
the corresponding principal $\Cal G$ bundle. The gauge algebra $Lie\Cal G$
can be identified with $\Omega ^0(M,adP)$ and the fundamental vector
field on $\Cal A$ with respect to the given $\Cal G$ action is
$Z_{\xi}=d_A\xi$
for $\xi\in Lie\Cal G$. Here $d_A\colon\Omega ^{\ast}(M ,adP)
\rightarrow\Omega ^{\ast +1}(M,adP)$ is the covariant exterior
derivative with respect to $A$. There is a natural scalar product on the
space $\Omega (M,adP)$ of all $adP$-valued differential forms on
$M$, given by
$$(\alpha,\beta ):=\int _{M}\
tr(\alpha\wedge\star\beta ),\qquad\alpha, \beta\in\Omega ^p(M,adP),
\tag2$$
where $\star$ is the corresponding Hodge operator, satisfying $\star ^2=(-1)
^{p(3-p)}$.\par
Let $i\colon\partial M\hookrightarrow M$ be the inclusion of the
boundary, then we denote the restriction of $P$ (i.e. pullback $i^{\ast}P$)
by $P_{\Sigma}:=i^{\ast}P$ with induced map $\bar i\colon
P_{\Sigma}\rightarrow P$. Let $\Cal B$ the space of all connections
on $P_{\Sigma}$ and let $\Cal H$ be the gauge group for $P_{
\Sigma}\rightarrow\Sigma$.
Evidently, $\bar i$ induces a map $\hat i\colon\Cal A\rightarrow
\Cal B$ by $\hat i(A)\equiv \hat A:=\bar i^{\ast}A$ and it also induces a map
$\Cal G\rightarrow\Cal H$. In the following we shall write $\hat\phi :=
\bar i^{\ast}\phi$ for the restriction of any $\phi\in\Omega ^{\ast}(M,
adP)$. If $\hat\star$ is the induced Hodge operator on $\Sigma$ satisfying
$\hat\star ^2=(-1)^{p(2-p)}$, an inner product on $\Omega ^{\ast}(\Sigma ,
adP_{\Sigma})$ is defined by
$(\phi _1,\phi _2)_{\Sigma}
=\int _{\Sigma}tr(\phi _1\wedge\hat\star\phi _2)$ for $\phi _i\in
\Omega ^p(\Sigma , adP_{\Sigma})$.
\par
Following Atiyah and Singer [9], let us consider the principal $G$ bundle
$\Cal B\times P_{\Sigma}@>>>\Cal B\times\Sigma$, which admits a natural
$\Cal H$ action. If the quotient is taken along this action, one obtains a
principal $G$ bundle (the so called universal bundle)
$\Cal B\times _{\Cal H}P_{\Sigma}@>>>
\Cal N\times\Sigma$,
where $\Cal N:=\Cal B/\Cal H$ denotes the
corresponding orbit space. There is a natural connection in the principal
$\Cal H$ bundle $\Cal B(\Cal N,\Cal H)$, namely
$$\alpha _B(\rho _B)=\hat G_B\hat d_B^{\ast}\rho _B,\qquad\rho _B
\in T_B\Cal B\cong\Omega ^1(\Sigma ,adP_{\Sigma}).\tag3$$
Here $\hat G_B=(\hat d_B^{\ast}\hat d_B)^{-1}$ is the Green
operator and $\hat d_B^{\ast}=-
\hat\star \hat d_B\hat\star$ is the adjoint of the covariant exterior
derivative $\hat d_B$ on $P_{\Sigma}$. Let us consider the following
connection
$$\omega _{(B,q)}(\rho _B,\zeta _q)=(\alpha _B(\rho _B))(q)+B_q(\zeta _q),
\qquad (\rho _B,\zeta _q)\in T_{(B,q)}(\Cal B\times P_{\Sigma})
\tag4$$
on $\Cal B\times P_{\Sigma}\rightarrow\Cal B\times\Sigma$,
which descends to a connection $\bar\omega$ on $\Cal B\times _{
\Cal H}P_{\Sigma}\rightarrow\Cal N\times\Sigma$.
According to the bigraded structure of the space of differential forms on
$\Cal B\times P_{\Sigma}$, the curvature $\Omega _{\omega}$ of $\omega$ is
determined by the components
$$\split &\Omega _{\omega\ (B,q)}^{\ \ (2,0)}(\rho _1,\rho _2) =
\Cal F_B(\rho _1,\rho _2)=\hat G_B\hat\star
([\rho _1^h,\hat\star \rho _2^h]-[\rho _2^h,\hat\star \rho _1^h])\\ &
\Omega _{\omega\ (B,q)}^{\ \ (1,1)}(\rho ,\zeta _q) =\rho _q^h(\zeta _q)\\ &
\Omega _{\omega\ (B,q)}^{\ \ (0,2)}(\zeta _q^1,\zeta _q^2)=F_B(\zeta _q^1,
\zeta _q^2),\endsplit\tag5$$
where $F_B$ is the curvature of $B$ and $\rho _i^h=\rho _i-d_B\alpha _B(
\rho _i)$, with $\rho _i\in T_B\Cal B$, are the horizontal projections
with respect to $\alpha$.\par
Another natural connection on $\Cal B\times P_{\Sigma}
\rightarrow\Cal B\times\Sigma$ is given by
$\eta _{(B,q)}(\rho _B,\zeta _q)=B_q(\zeta _q)$,
which, however, does not descend to the $\Cal H$ quotient. For $\rho _B\in
T_B\Cal B$ and $\zeta _q^i\in T_qP_{\Sigma}$ the components of its
curvature $\Omega _{\eta}$ are given by
$$\Omega _{\eta }^{\ \ (2,0)}=0,\qquad\Omega _{\eta\ (B,q)}
^{\ \ (1,1)}(\rho _B,\zeta _q)=(\rho _B)_q(\zeta _q),\qquad
\Omega _{\eta\ (B,q)}^{\ \ (0,2)}(\zeta _q^1,
\zeta _q^2)=F_B(\zeta _q^1,\zeta _q^2).\tag6$$
Analogously, we can define a connection $\varphi$
in the principal $G$ bundle $\Cal A\times P\rightarrow\Cal A
\times M$ by
$\varphi _{(A,p)}(\tau _A,X_p):=A_p(X_p)$,
whose curvature $\Omega _{\varphi}$ has the following components
$$\Omega _{\varphi }^{\ \ (2,0)}=0,\qquad\Omega _{\varphi\ (A,p)}
^{\ \ (1,1)}(\tau _A,
X_p)=(\tau _A)_p(X_p),\qquad\Omega _{\varphi\ (A,p)}^{\ \ (0,2)}
(X_p^1,X_p^2)=F_A(X_p^1,X_p^2),\tag7$$
where $\tau _A\in T_A\Cal A$ and $X_p^i\in T_pP$. It is evident that
$(id\times\bar i)^{\ast}\varphi =(\hat i\times id)^{\ast}\eta$ holds.
\par
In the remainder of this paper we shall consider the trace as an
ad-invariant polynomial $Q$ on $\frak g$ of degree 2.
Let $\Cal P @>>> X$ be any of the principal bundles introduced before
and let $\alpha$ be a corresponding
connection on $\Cal P$ with curvature $F$. Then the exact
4-form $Q(F):=Q(F,F)$ on $\Cal P$ descends to a well defined form on $X$.
If $\alpha _1$ and $\alpha _2$ are two connections on $\Cal P$ with
curvatures
$F_1$ and $F_2$, then the secondary characteristic 3-form
$TQ(\alpha _1,\alpha _2)\in\Omega ^3(X,\Bbb R)$ satisfies [12]
$$Q(F_1)-Q(F_2)=d_{X}\ TQ(\alpha _1,\alpha _2),\tag8$$
where $TQ(\alpha _1,\alpha _2)=2\int _0^1dt\ Q(\alpha _1-\alpha _2,\Cal F_t)$
and $\Cal F_t$ is the curvature of the interpolating connection $(1-t)
\alpha _2+t\alpha _1$.
\par
If $G$ is not simply connected, then the bundles $P$ and
$P_{\Sigma}$ may be nontrivial. In that case we choose a fixed
background field $a\in\Cal A$ which is
extended to a connection in $\Cal A\times P$ in a natural way (we shall
denote it with the same symbol) and has
curvature $\Omega _a$. Because of dimensional reasons, $Q(\Omega _a)=0$.
Let $\hat a$ denote the restriction of this connection to $\Cal A\times P_{
\Sigma}$.\par
The Chern-Weil formula (8) yields the following set of descent equations
for the connections $\omega$, $\eta$, $\varphi$ and $a$
$$\align & d_{\Cal B}Q(\Omega _{\omega})^{(k,4-k)}+(-1)^{k+1}d_{
\Sigma}Q(\Omega _{\omega})^{(k+1,3-k)}=0\tag"(9a)"\\ &
d_{\Cal B}Q(\Omega _{\eta})^{(k,4-k)} +(-1)^{k+1}d_{\Sigma}
Q(\Omega _{\eta})^{(k+1,3-k)}=0\tag"(9b)"\\ &
d_{\Cal A}Q(\Omega _{\varphi})^{(k,4-k)} +(-1)^{k+1}d_{M}
Q(\Omega _{\varphi})^{(k+1,3-k)}=0\tag"(9c)"\\ &
Q(\Omega _{\omega})^{(k,4-k)}=d_{\Cal B}TQ(\omega ,\hat a)^{(k-1,4-k)} +
(-1)^{k}d_{\Sigma}TQ(\omega ,\hat a)^{(k,3-k)}\tag"(9d)"\\
& Q(\Omega _{\eta})^{(k,4-k)}=d_{\Cal B}TQ(\eta ,\hat a)^{(k-1,4-k)} +
(-1)^{k}d_{\Sigma}TQ(\eta ,\hat a)^{(k,3-k)}\tag"(9e)"\\
& Q(\Omega _{\varphi})^{(k,4-k)}=d_{\Cal A}TQ(\varphi ,a)^{(k-1,4-k)} +
(-1)^{k}d_{M}TQ(\varphi ,a)^{(k,3-k)}\tag"(9f)"\\ &
(Q(\Omega _{\omega})-Q(\Omega _{\eta}))^{(k,4-k)}=d_{\Cal B}TQ(\omega ,
\eta)^{(k-1,4-k)} +
(-1)^{k}d_{\Sigma}TQ(\omega ,\eta)^{(k,3-k)},\tag"(9g)"
\endalign$$
where $d_{\Cal A}$, $d_{\Cal B}$, $d_M$, $d_{\Sigma}$ are the corresponding
exterior derivatives and the superscripts indicate the form degrees
with respect to the bigraded structure of the algebra of differential
forms on $\Cal A\times M$ and $\Cal B\times\Sigma$, respectively.
Finally, the secondary characteristic forms satisfy the following identity
$$TQ(\omega ,\eta )=TQ(\omega ,\hat a)-TQ(\eta ,\hat a)+d_{\Cal A\times
\Sigma}\ SQ(\omega ,\eta ,\hat a),\tag10$$
for the 2-form $SQ(\omega ,\eta ,\hat a)=Q(\omega -\eta ,\eta -
\hat a)\in\Omega ^2(\Cal B\times \Sigma,\Bbb R)$.
An application of these descent equations for the determination of covariant
Yang-Mills anomalies has been discussed in [10].
\bigskip\bigskip
{\bf  3. The classical phase space of the PYM theory}
\bigskip
In this section we want to analyze the classical structure of the model
defined by the action (1). We leave both the values of the gauge fields
and the gauge transformations free on the boundary.
Since we want to study the system in a fixed time formalism, we introduce the
space $\Cal A^0=\Omega ^0(M ,adP)$ of all scalar potentials. The
classical
configuration space is $\Cal A\times\Cal A^0$ and the Lagrangian associated
to (1) is the real valued function
$$L(A,A_0,\dot A,\dot A_0)=\frac{1}{2}\Vert \dot A-d_{A}A_0\Vert ^2
-\frac{1}{2}\Vert F_A\Vert ^2
+\frac{\theta}{8\pi ^2 }(\dot A-d_AA_0,\star F_A),\tag11$$
where $\Vert\cdot\Vert$ is the norm associated with (2) and
$(\dot A,\dot A_0)\in\Omega ^1(M,adP)\times\Omega ^0(M,
adP)$ are the fiber coordinates in the tangent bundle $T(\Cal A\times
\Cal A^0)$.
\par
Elements of the phase space are
pairs $(A,A_0,\Pi ,\Pi _0)$ where, according to the inner product (2),
$(\Pi ,\Pi _0)$ is regarded as an element
of $\Omega ^1(M,adP)\times\Omega ^0(M,adP)$. The conjugate momenta are
given by
$$\Pi _0=0,\qquad\Pi =\frac{\delta L}{\delta A}=\dot A-d_{A}A_0+\frac{
\theta}{8\pi ^2}\star F_A.\tag12$$
Hence the corresponding Hamilton function reads
$$H(A,A_0,\Pi ,\Pi _0)=\frac{1}{2}\Vert \Pi -\frac{\theta}{8\pi ^2}
\star F_A\Vert ^2+\frac{1}{2}\Vert F_A\Vert ^2+(\Pi ,d_AA_0).\tag13$$
The Poisson bracket $\lbrace ,\rbrace$ is induced by the canonical
symplectic two form on $T^{\ast}(\Cal A\times\Cal A_0)$
$$\multline \frak K_{\Cal A}((\tau _1,\tau _1^0,\sigma _1 ,\sigma _1^0),
(\tau _2,\tau _2^0,\sigma _2 ,\sigma _2^0))\\ =
\int _{M}\left(Q(\tau _2,\star\sigma _1)+
Q(\tau _2^0,\star\sigma _1^0)-Q(\tau _1,\star\sigma _2)-Q(\sigma _1^0,\star
\tau _2^0)\right),\endmultline\tag14$$
with
$(\tau _i,\tau _i^0,\sigma _i ,\sigma _i^0)
\in T(T^{\ast}(\Cal A\times\Cal A_0))$, $i=1,2$.
The Lagrangian is singular and gives rise to the primary constraint
$\Pi _0=0$. The secondary constraint is given by
$$J_{\xi}=\lbrace H,\Pi _0(\xi )\rbrace =(\Pi ,d_A\xi ),\qquad\xi\in Lie
\Cal G\tag15$$
which is of first class $\lbrace J_{\xi},J_{\eta}\rbrace =J_{[\xi ,\eta ]}$,
and there are no constraints of higher order, since
$\lbrace J_{\xi},H\rbrace =J_{[A_0,\xi ]}$.
In the following we eliminate the constraint $\Pi _0=0$ by fixing $A_0=0$.
\par
We want to note that the constraint algebra of the PYM model is of first
class whereas it is of second class in the CSYM theory on a manifold with
boundary. In the latter only gauge
transformations which reduce to the identity at the boundary can be regarded
as gauge degrees of freedom (see Ref. 2 and references therein).
\par
In order to identify the reduced classical phase space of our model, we note
that the
Hamiltonian (13) describes the motion of the Yang-Mills field in the
background of
the abelian functional field $\frac{\theta}{8\pi ^2}\star F_A$.
Instead of
analyzing the theory on $T^{\ast}\Cal A$ with Hamiltonian (13) and
symplectic structure (14), we follow the method proposed by
Sternberg [13]
to rewrite this system in terms of the new momenta $\Pi -\frac{\theta}
{8\pi ^2}\star F_A$. In consequence, this modifies the canonical symplectic
structure.
Therefore, let us consider the following diffeomorphism $\gamma$
along the
fibers of the cotangent bundle $T^{\ast}\Cal A @>\pi _T>> \Cal A$
$$\gamma (\Pi )_A(\tau ):=(\Pi ,\tau )+\frac{\theta}{8\pi ^2}(\tau ,
\star F_A)=(\Pi ,\tau )+\frac{\theta}{16\pi ^2}\int _{M}Q(\Omega _
{\varphi})^{(1,3)}(\tau ),\tag16$$
where $\Pi\in T_A^{\ast}\Cal A$ and $\tau\in T_A\Cal A$. For the
identification of the last term on the right hand side of (16) we have used
(7).\par
Setting $\beta :=\frac{\theta}{16\pi ^2}\int _{M}Q(\Omega _
{\varphi})^{(1,3)}$ and
using (9c), the new symplectic structure which
includes the interaction with the functional background field reads
$$\frak K_{\beta}:=\gamma ^{\ast}\frak K_{\Cal A}=
\frak K_{\Cal A} +\pi _{T}^{\ast}\ R_{\beta},\tag17$$
where the two form $R_{\beta}:=d_{\Cal A}\beta\in\Omega ^2(\Cal A,\Bbb R)$
is given by
$$R_{\beta}(\tau _1,\tau _2)=-\frac{\theta}{16\pi ^2}\ \hat i^{\ast}
\int _{\Sigma}Q(\Omega_{\eta})^{(2,2)}(\tau _1,\tau _2)=
\frac{\theta}{8\pi ^2}\int _{\Sigma}Q(\hat\tau _1,\hat\tau _2),\qquad
\tau _i\in T_A\Cal A.\tag18$$
Finally, the new Hamiltonian has the form
$$\tilde H=\frac{1}{2}\Vert \Pi\Vert ^2+\frac{1}{2}\Vert F_A\Vert ^2+
(\Pi +\frac{\theta}{8\pi ^2}\star F_A ,d_AA_0),\tag19$$
and therefore
$$\tilde J_{\xi}(A,\Pi )=(\Pi ,d_A\xi )+\frac{\theta}{8\pi ^2}\int _{
\Sigma}Q(\hat\xi ,F_{\hat A})\tag20$$
is the corresponding Gauss-law constraint.
Note that the second term in (20) coincides with the expression for the
two dimensional covariant anomaly [10] on $\Sigma$. \par
There is a natural symplectic action of the gauge group $\Cal G$ on
$(T^{\ast}\Cal A,\frak K_{\beta})$ with infinitesimal generator
$X_{\xi}\in\frak X(T^{\ast}\Cal A)$,
$$X_{\xi}(\Phi )(A,\Pi )=\frac{d}{dt}\vert _{t=0}\left( \Phi (A+td_A\xi ,
\Pi )-\Phi (A,\Pi +t[\xi ,\Pi ])\right),\qquad\Phi\in
C^{\infty}(T^{\ast}\Cal A),\tag21$$
satisfying $i_{X_{\xi}}\frak K_{\beta}
=-d_{T^{\ast}\Cal A}\ \tilde J_{\xi}$.\par
The classical PYM model is described by the constrained system
$(T^{\ast}\Cal A,\frak K_{\beta},\tilde J)$ consisting of the symplectic
manifold $(T^{\ast}\Cal A,\frak K_{\beta})$, the Gauss constraint
$\tilde J$ which is viewed as an equivariant momentum map
$T^{\ast}\Cal A @>>> (Lie\Cal G)^{\ast}$ from the phase space to the dual of
the gauge algebra and the Hamiltonian $\tilde H$.
Via the Marsden Weinstein reduction [14],
the true phase space of the model is obtained as the quotient
$\tilde J^{-1}(0)/\Cal G$ and the symplectic form $\overline{\frak K_{
\beta}}$ is given
by restricting $\frak K_{\beta}$ to $\tilde J^{-1}(0)$ and projecting onto
the orbit space.
\proclaim{Proposition 1} Consider the constrained system $(T^{\ast}\Cal A,
\frak K_{\beta},\tilde J)$. There exists a symplectomorphism between the
symplectic manifolds
$(\tilde J^{-1}(0)/\Cal G,\overline{\frak K_{\beta}})$ and
$(T^{\ast}\Cal M,\overline{\frak K_{\omega}})$.\endproclaim
\demo{Proof} Let us consider the symplectic $\Cal G$ space $(T^{\ast}\Cal A,
\frak K_{\omega},J)$, with symplectic two form
$\frak K_{\omega}=\frak K_{\Cal A}+\pi _{T}^{\ast}R_{\omega}$, where
$$R_{\omega}(\tau _1,\tau _2)=-\frac{\theta}{16\pi ^2}\hat i^{\ast}
\int _{\Sigma}Q(\Omega _{\omega})^{(2,2)}(\tau _1,\tau _2)=
\frac{\theta}{8\pi ^2}\int _{\Sigma}[Q(\hat\tau _1^h,\hat\tau _2^h)-
Q(\Cal F _{\hat A}(\hat\tau _1,\hat\tau _2),F_{\hat A})],\tag22$$
and $J_{\xi}=(\Pi ,d_A\xi )$ (15) is the equivariant momentum
satisfying $i_{X_{\xi}}\frak K_{\omega}
=-d_{T^{\ast}\Cal A}\ J_{\xi}$.\par
The two form $R_{\omega}$ descends to $\overline{R_{\omega}}$, which is
obtained by pulling back $-\frac{\theta}
{16\pi ^2}\int _{\Sigma}Q(\Omega _{\bar\omega})^{(2,2)}$ along the induced
map $\Cal M @>>>\Cal N$.
Let $\bar\pi _{T}\colon T^{\ast}\Cal M\rightarrow\Cal M$ be the projection
and let $\frak K_{\Cal M}$ be the canonical symplectic form on $T^{\ast}
\Cal M$ then the symplectic reduction of $(T^{\ast}\Cal A,
\frak K_{\beta},\tilde J)$ yields $(T^{\ast}\Cal M,\overline{
\frak K_{\omega}})$,
where $\overline{\frak K_{\omega}}=\frak K_{\Cal M}+\bar\pi _{T}^{\ast}
\overline{R_{\omega}}$.\par
Let $j:J^{-1}(0)\hookrightarrow T^{\ast}\Cal A$ and
$\tilde j\colon
\tilde J^{-1}(0)\hookrightarrow T^{\ast}\Cal A$ be the inclusions, then we
consider the following diffeomorphism along the fibers of $T^{\ast}\Cal A
\rightarrow\Cal A$
$$\chi (\Pi )_A(\tau ):=(\Pi ,\tau )-\frac{\theta}{16\pi ^2}
\int _{\Sigma}TQ(\omega ,\eta )^{(1,2)}(\hat\tau )=(\Pi ,\tau )-
\frac{\theta}{8
\pi ^2}(\hat d_{\hat A}\hat G_{\hat A}\hat\star F_{\hat A},
\hat\tau )_{\Sigma},\tag23$$
where $\tau\in T_A\Cal A$ and the explicit expression for $TQ(\omega ,\eta )$
was calculated using (5) and (6). \bigskip
Because of (9g) one obtains $\frak K_{\omega}=\chi ^{\ast}
\frak K_{\beta}$, so that $\chi$ induces a diffeomorphism $J^{-1}(0)
\rightarrow \tilde J^{-1}(0)$, which also is $\Cal G$ equivariant.
Finally, the relation $\tilde j\circ\chi =\chi\circ j$ implies that $\chi$
is a presymplectomorphism.
\qed\enddemo
In the new coordinates, the dynamical structure of the model is thus governed
by the Hamiltonian
$$H_{\theta}= \frac{1}{2}\Vert \Pi\Vert ^2
-\frac{\theta}{8\pi ^2}(\hat\Pi ,\hat d_{\hat A}\hat G_{\hat A}\hat\star F_{
\hat A})_{\Sigma}+\frac{1}{2}\Vert F_A\Vert ^2 +(\Pi ,d_AA_0)
+\frac{1}{2}(\frac{
\theta}{8\pi ^2})^2\Vert \hat d_{\hat A}\hat G_{\hat A}\hat\star F_{\hat A}
\Vert _{\Sigma}^2,\tag24$$
which descends to a Hamiltonian $\bar H_{\theta}$ on the symplectic quotient
$T^{\ast}\Cal M$.
We should remark that the symplectomorphism between
$\tilde J^{-1}(0)/\Cal G$
and $T^{\ast}\Cal M$ requires the introduction of a connection in
$\Cal B(\Cal N,\Cal H)$. However, with our choice (3), the symplectic data
of the PYM model are similar to those of the CSYM theory [5]. This is a
consequence of the fact that the Pontryagin density is the derivative
of the Chern-Simons term.
\bigskip\bigskip
{\bf 4. Quantization of the PYM theory}
\bigskip
In this section we want to study the quantum theory of the PYM model in the
framework of geometric quantization [10]. Essentially, one has to introduce
a prequantum line bundle and a polarization of the phase space.
We consider the extended phase
space quantization of $(T^{\ast}\Cal A,\frak K_{\omega},J)$
following Dirac [15]. The idea is to quantize the
unconstrained
system and then to impose the constraints as conditions on the states.
\par
Let $\Cal L_0^{\prime}=T^{\ast}\Cal A\times\Bbb C$ be the trivial
prequantum line bundle over $T^{\ast}\Cal A$
with connection $\nabla :=d_{T^{\ast}\Cal A}-i\vartheta -i\varepsilon$, where
$$\varepsilon =-\frac{\theta}{16\pi ^2}\ \hat i^{\ast}\int _{\Sigma}TQ(
\omega ,\hat a)^{(1,2)}\in\Omega ^1(\Cal A,\Bbb R).\tag25$$
It is evident from (9d) and (22) that $d_{\Cal A}\varepsilon =R_{\omega}$.
Associated with any observable $\Phi\in C^{\infty}(T^{\ast}\Cal A)$
is the first order differential operator
$\Cal O_{\Phi}:=-i\nabla _{X_{\Phi}}+\Phi$,
where $X_{\Phi}$ is defined by $i_{X_{\Phi}}\frak K_{\omega}=-
d_{T^{\ast}\Cal A}\ \Phi$. Let $\Cal O_{\xi}$ denote the Gauss constraint
operator associated with the momentum map (15) $J_{\xi}
\in C^{\infty}(T^{\ast}\Cal A)$, then the physical admissible states are
those sections of $\Cal L_0^{\prime}$ which are annihilated by
$\Cal O_{\xi}$ and compatible with the chosen polarization.
In the Schr\"odinger polarization of $T^{\ast}\Cal A$, this requires the
restriction to sections $\psi$ of the trivial line bundle $\Cal L_0$ on
$\Cal A$, satisfying
$$\nabla _{Z_{\xi}}^{
\varepsilon}\psi (A)=L_{Z_{\xi}}\psi (A)+\frac{i\theta}{16\pi ^2}(\hat\xi ,
\hat\star (F_{\hat A}+F_{\hat a}-\frac{1}{2}[\hat A-\hat a,\hat A-\hat a]))_{
\Sigma}\ \psi (A)=0,\qquad\xi\in Lie\Cal G\tag26$$
where $\nabla ^{\varepsilon}:=d_{\Cal A}-i\varepsilon$ is the covariant
derivative on $\Cal L_0$ with curvature $-iR_{\omega}$ and
$L_{Z_{\xi}}$ is the Lie derivative on $\Cal A$ along the fundamental vector
field $Z_{\xi}$.
The conjugate momentum is represented by the operator
$-i\nabla ^{\varepsilon}$ and hence the Hamilton operator reads
$$H_{\theta}= -\frac{1}{2}\Vert \nabla ^{\varepsilon}\Vert ^2
+\frac{i\theta}{8\pi ^2}(\hat\nabla ^{\varepsilon} ,\hat d_{\hat A}\hat
G_{\hat A}\hat\star F_{\hat A})_{
\Sigma}+\frac{1}{2}\Vert F_A\Vert ^2
+\frac{1}{2}(\frac{
\theta}{8\pi ^2})^2\Vert \hat d_{\hat A}\hat G_{\hat A}\hat\star F_{\hat A}
\Vert _{\Sigma}^2.\tag27$$
Notice that the second term on the right hand side of
(27) is the expression for the operator $\Cal O_f$, evaluated in
the Schr\"odinger polarization, where
$f:=-\frac{\theta}{8\pi ^2}
(\hat\Pi ,\hat d_{\hat A}\hat G_{\hat A}\hat
\star F_{\hat A})_{\Sigma}$ in (24) is regarded as pullback of a function
in $C^{\infty}(T^{\ast}\Cal B)$. \par
In order to solve the Gauss constraint (26), we shall
comment on its geometrical meaning.
Since $\Cal G$ is not necessarily connected for a general principal
$G$-bundle $P$ on the 3-manifold $M$, only
the behaviour of physical states under gauge transformations
belonging to the connected component of the identity $\Cal G_0$ of $\Cal G$
is controlled by the Gauss constraint. Here the gauge group $\Cal G$
appears in the exact sequence
$$1\rightarrow\Cal G_0\rightarrow\Cal G\rightarrow\pi _0(\Cal G)\rightarrow
1,\tag28$$
where $\pi _0(\Cal G)$ denotes the group of components of $\Cal G$.
Correspondingly, one has the principal $\Cal G_0$ bundle $\Cal A@>
\pi ^{\prime}
>>\tilde\Cal M=\Cal A/\Cal G_0$, where $\tilde \Cal M@>\tilde\pi >>\Cal M$
itself has the structure of a principal $\pi _0(\Cal G)$ bundle over the
gauge orbit space $\Cal M$. Since
$\pi _1(\Cal M)=\pi _0(\Cal G)$, $\tilde\Cal M$
is the simply connected covering space of $\Cal M$. \par
There is a natural $Lie\Cal G$ action on $\Cal L_0$ induced by the
horizontally (with respect to $\nabla ^{\varepsilon}$) lifted fundamental
vector fields $Z_{\xi}$. If the restrictions of $\Cal L_0$
along the $\Cal G_0$ orbits in $\Cal A$ have trivial holonomy, the
infinitesimal action can be extended to an action of $\Cal G_0$ on
$\Cal L_0$. This obstruction is a manifestation of the quantization condition
in the Dirac approach.\par
Let $j_A\colon\Cal G\rightarrow\Cal A$, $j_A(u)=u^{\ast}A$ be the natural
inclusion of the fiber then we define the functional
$F^{\prime}(A,u,c):=exp(2\pi i\int _cj_A^{\ast}\varepsilon )$ for $(A,u)\in
\Cal A\times\Cal G_0$ where $c$ is a
path in $\Cal G_0$ with $c(0)=id_{\Cal G}$ and $c(1)=u$. It is evident that
the
functional $F^{\prime}$ is independent of the choice of the path $c$ if and
only if the cohomology class $[j_A^{\ast}\varepsilon ]$ belongs to
$H^1(\Cal G_0,\Bbb Z)$ which implies that the holonomy of $j_A^{\ast}
\Cal L_0$ is trivial.
In that case we write $F(A,u)=F^{\prime}(A,u,c)$
and the $\Cal G_0$ action on $\Cal L_0$ is given by $(A,z)\mapsto (u^{\ast}
A,F(A,u)z)$. This action would be well defined since
$F(u_1^{\ast}A,u_2)F(A,u_1)=F(A,u_1u_2)$. \par
In order to show that the quantization obstruction is trivially fulfilled in
the present case, let us consider the following one form on $\Cal A$
$$\kappa =\frac{\theta}{16\pi ^2}\left( \int _{M}Q(\Omega _{
\varphi})^{(1,3)}-
\int _{\Sigma}(\hat i\times id)^{\ast}TQ(\omega ,\eta )^{(1,2)}
\right).\tag29$$
Using (9c,g) it is easy to prove that $R_{\omega}=d_{\Cal A}
\kappa$.
Since $Q(\Omega _{\omega})^{(2,2)}$ is horizontal, one concludes from (7)
and (10)
that $i_{Z_{\xi}}\kappa =0$ and hence $\kappa$ descends to a form
$\bar\kappa
\in\Omega ^1(\Cal M,\Bbb R)$ such that $\overline{R_{\omega}}=d_{\Cal M}\bar
\kappa$. Hence we have proven
\proclaim{Proposition 2} The De Rham class of $\overline{R_{\omega}}$ in
$H^2(\Cal M,\Bbb R)$ is trivial.\endproclaim
It follows from (10), (25) and (29) that
$$j_A^{\ast}\varepsilon =-d_{\Cal G_0}\ j_A^{\ast}\ \frac{\theta}{16\pi ^2}
\int _{M}TQ(\varphi ,a)^{(0,3)}\tag30$$
is an exact one form on $\Cal G_0$ and since $TQ(\varphi ,a)^{(0,3)}=
TQ(A,a)$, one obtains
$$F(A,u)=exp\left( -\frac{i\theta}{8\pi}(\int _{M}TQ(u^{\ast}A,a)-
\int _{M}TQ(A,a))\right) .\tag31$$
If we factorize by this $\Cal G_0$ action, we get the line bundle
$\Cal L_F=\Cal A\times _F\Bbb C$ on $\tilde\Cal M$.
Let $\gamma (t)$ be a path in $\Cal A$ with $\gamma (0)=A$ and
$u\in\Cal G_0$ then $\varepsilon$ satisfies
$$\varepsilon _{u^{\ast}A}(\frac{d}{dt}\vert _{t=0}\ u^{\ast}\gamma (t))-
\varepsilon _{A}
(\frac{d}{dt}\vert _{t=0}\ \gamma (t))=\frac{d}{dt}
\vert _{t=0}\ F(A,u)^{-1}F(\gamma (t),u),\tag32$$
which is the necessary and sufficient condition for $\varepsilon$
descending to a well defined connection $\bar\varepsilon$ on the line
bundle $\Cal L_F$.
Since ($\Cal L_F,\bar\varepsilon )$ is the unique (up to bundle
equivalence)
line bundle on $\tilde\Cal M$ with curvature $-i\tilde\pi ^{\ast}
\overline{R_{\omega}}$, it is trivializable and therefore no quantization
condition
must be imposed on the coupling parameter $\theta$ to obtain non-trivial
solutions of (26).
\par
Let $Aut(\Cal L_F,\bar\varepsilon )$ denote the group of bundle
automorphisms
on $\Cal L_F$ leaving the connection $\bar\varepsilon$ invariant and let
$r\colon\tilde\Cal M\times \pi _0(\Cal G)\rightarrow\tilde\Cal M$ be the
induced principal action of $\pi _0(\Cal G)$ on $\tilde\Cal M$.
Since $\Cal L_F$ is trivializable, there exists a lifting [16]
$\nu\colon\pi _0(\Cal G)\rightarrow
Aut(\Cal L_F,\bar\varepsilon)$ of $r$. The space of orbits
$\Cal L_{F,\nu}:=\Cal L_F/
\nu$ of $\nu$ on $\Cal L_F$ is a line bundle on $\Cal M$ with induced
connection
$\tilde\varepsilon$ and curvature $-i\overline{R_{\omega}}\in\Omega ^2(
\Cal M,\Bbb R)$.
\par
According to Prop. 2, the quantum line bundle $\Cal L_{F,\nu}$ has
vanishing real Chern class and therefore its topological type
is classified by the image of the Bockstein operator
$\delta ^{\ast}\colon H^1(\Cal M,\Bbb R/\Bbb Z) @>>>H^2(\Cal M,\Bbb Z)$.
Notice that $H^1(\Cal M,\Bbb R/\Bbb Z)\cong Hom(\pi _0(\Cal G),\Bbb R/
\Bbb Z)$.
In fact, $\Cal L_{F,\nu}$ also depends on the chosen background field $a\in
\Cal A$ but it
can be easily shown that the line bundles corresponding to different
background fields are isomorphic.
\proclaim{Proposition 3} The space of physical admissible states is
isomorphic to the
space of sections of the line bundle $\Cal L_{F,\nu}$ on $\Cal M$.
\endproclaim
Although the wave functionals can be chosen to be $\Cal G_0$ invariant,
a non-trivial lift $\nu$
represents an obstruction to extend them to the whole
$\Cal A$ in a $\Cal G$ invariant way.
If the abelization of $\pi _1(\Cal M)=\pi _0(\Cal G)$, namely $H_1(\Cal M,
\Bbb Z)$, is torsionless, then $\Cal L_{F,\nu}$ will be trivializable.
\par
Let us remark that for suitable manifolds the two form
$\int _{\Sigma}Q(\Omega _{\bar\omega})^{(2,2)}$ belongs to
the generating class in $H^2(\Cal N,\Bbb Z)$ [6]. Hence the $\theta$
parameter
would be quantized if and only if the quantum line bundle $\Cal L_{F,
\Cal\nu}$ was the
pullback of a line bundle on $\Cal N$. However, according to the
principles of geometric
quantization, there is no reason that this may be the case.\par
The isomorphism class of $\Cal L_{F,\nu}$ (as a unitary line bundle with
connection) is not unique, giving rise to inequivalent quantum theories,
because the lift $\nu$ is determined only up to
elements of $Hom(\pi _0(\Cal G),\Bbb R/\Bbb Z)$.
Generally, the possible quantum line bundles are of the form
$\Cal L_{F,\nu}\otimes\Cal V$,
where $\Cal V$ is a flat unitary line bundle on $\Cal M$.
\bigskip\bigskip\bigskip
{\bf References}
\bigskip
\roster
\item"{[1]}" R. Jackiw and S. Templeton, Phys. Rev. {\bf D 23} (1981) 2291;
J. Schonfeld, Nucl. Phys. {\bf B 185} (1981) 157;
S. Deser, R. Jackiw and S. Templeton, Phys. Rev. Lett. {\bf 48}
(1982) 975; Ann. Phys. {\bf 140} (1982) 372.
\item"{[2]}" G.V. Dunne and C.A. Trugenberger, Ann. Phys. {\bf 204}
(1990) 281.
\item"{[3]}" A.A. Belavin, A.M. Polyakov, A.S. Schwarz and Yu. Tyupkin, Phys.
Lett. {\bf B 59} (1975) 85;
R. Jackiw and C. Rebbi, Phys. Rev. Lett. {\bf 37} (1976) 172;
C. Callan, R. Dashen and D. Gross, Phys. Lett. {\bf B 63} (1976) 334;
G.t Hooft, Phys. Rev. {\bf D 37} (1976) 8; Phys. Rev. {\bf D 14}
(1976) 334.
\item"{[4]}" Y.S. Wu and A. Zee, Nucl. Phys. {\bf B 258} (1985) 157.
\item"{[5]}" M. Asorey and P.K. Mitter, Phys. Lett. {\bf B 153} (1985) 147.
\item"{[6]}" M. Asorey and P.K. Mitter, Ann. Inst. H. Poincare {\bf 45}
(1986) 61.
\item"{[7]}" J. Mickelsson, Lett. Math. Phys. {\bf 7} (1983) 45; Commun.
Math. Phys. {\bf 97} (1985) 361; L. Faddeev, Phys. Lett. {\bf B 145}
(1984) 91.
\item"{[8]}" E. Witten, Commun. Math. Phys. {\bf 121} (1989) 351; A.P.
Balachandran, G. Bimonte, K.S. Gupta and A. Stern, Int. J. Mod. Phys. {\bf A
7} (1992) 4655.
\item"{[9]}" M.F. Atiyah und I.M. Singer, Proc. Natl. Acad. Sci. {\bf 81}
(1984) 2597.
\item"{[10]}" G. Kelnhofer, J. Math. Phys. {\bf 34} (1993) 3901.
\item"{[11]}" N.M.J. Woodhouse, Geometric quantization (2nd ed., Clarendon:
Oxford, 1992)
\item"{[12]}" S.S. Chern and J. Simons, Ann. Math. {\bf 99} (1974) 48.
\item"{[13]}" S. Sternberg, Proc. Natl. Acad. Sci. {\bf 74} (1977) 5253.
\item"{[14]}" J.E. Marsden and A. Weinstein, Rep. Math. Phys. {\bf 5}
(1974) 121.
\item"{[15]}" P.A.M. Dirac, Lectures on Quantum Mechanics (Academic,
New York, 1964).
\item"{[16]}" B. Kostant, Springer Lecture Notes in Mathematics 170
(1970) 87.

\endroster

\enddocument